\documentclass[useAMS,usenatbib]{mn2e}
\usepackage{txfonts}
\usepackage{natbib}
\usepackage[all]{xy}
\usepackage[dvips]{graphicx}

\def\del#1{{}}

\newcommand{\ltsima}{$\; \buildrel < \over \sim \;$}
\newcommand{\lsim}{\lower.5ex\hbox{\ltsima}}
\newcommand{\gtsima}{$\; \buildrel > \over \sim \;$}
\newcommand{\gsim}{\lower.5ex\hbox{\gtsima}}

\newcommand{\dd}{\mathrm{d}}

\title[iSW-effect in coupled models]
{The integrated Sachs-Wolfe effect in cosmologies with coupled dark matter and dark energy}
\author[Bj{\"o}rn Malte Sch{\"a}fer]
{Bj{\"o}rn Malte Sch\"afer\thanks{e-mail: bjoern.malte.schaefer@ias.u-psud.fr}\\
Institute of Cosmology and Gravitation, University of Portsmouth, Mercantile House, Hampshire Terrace, Portsmouth PO1 2EG, United Kingdom\\
Institut d'Astrophysique Spatiale, Universit{\'e} de Paris XI, b{\^a}timent 120-121, Centre universitaire d'Orsay, 91400 Orsay CEDEX, France}

\begin{document}
\pagerange{\pageref{firstpage}--\pageref{lastpage}}
\pubyear{2008}
\maketitle
\label{firstpage}

\begin{abstract}
The subject of this paper is the derivation of the integrated Sachs-Wolfe (iSW) effect in cosmologies with coupled dark matter and dark energy fluids. These couplings influence the iSW-effect in three ways: The Hubble function assumes a different scaling, the structure growth rate shows a different time evolution, and in addition, the Poisson equation, which relates the density perturbations to fluctuations in the gravitational potential, is changed, due to the violation of the scaling $\rho\propto a^{-3}$ of the matter density $\rho$ with scale factor $a$. Expemplarily, I derive the iSW-spectra for a model in which dark matter decays into dark energy, investigate the influence of the dark matter decay rate and the dark energy equation of state on the iSW-signal, and discuss the analogies for gravitational lensing. Quite generally iSW-measurements should reach similar accuracy in determining the dark energy equation of state parameter and the coupling constant.
\end{abstract}


\section{Introduction}
Recently, cosmological models with coupling terms in the evolution equations for the dark matter and the dark energy density have attracted interest, because these couplings alleviate the coincidence problem in an elegant way \citep{2003PhRvD..67h3513C, 2007arXiv0706.3860O, 2008arXiv0802.1086P, 2008arXiv0801.1565B, 2008arXiv0801.4233H}.

The integrated Sachs-Wolfe (iSW) effect \citep{1967ApJ...147...73S, rees_sciama_orig, 1994PhRvD..50..627H, 2002PhRvD..65h3518C, 2006MNRAS.369..425S}, which refers to the frequency change of cosmic microwave background (CMB) photons if they cross time evolving gravitational potentials, is a direct probe of dark energy because it vanishes in cosmologies with $\Omega_m=1$ \citep{1996PhRvL..76..575C}. By now, it has been detected with high significance with a number of different tracer objects \citep{2003AIPC..666...67B, 2003ApJ...597L..89F, 2006PhRvD..74f3520G, 2007MNRAS.377.1085R, 2008arXiv0801.4380G}.

In this paper, I would like to focus on the derivation of the integrated Sachs-Wolfe effect for cosmological models with coupled dark matter and dark energy fluids and to show that iSW-effect acquires a dependence on the Hubble function, the matter density parameter, the growth rate and their derivatives. In this respect, the iSW-effect in coupled models is much richer than in models with non-interacting constituents, where the iSW-effect simply measures the line of sight integrated derivative $\dd (D_+/a)/\dd a$ of the growth function $D_+(a)$. Physically, the changes in the iSW-formulae are due to the fact that the matter density $\rho_m$ does not develop $\propto a^{-3}$ in coupled models, and the relation $\Omega_m(a)/\Omega_m=H_0^2/H^2(a)/a^3$ does not hold. A consequence of this is a different (time-evolving) relation between the gravitational potential and the overdensity field; and it is not clear how mechanism was incorporated in the derivation presented by \citet{2008arXiv0801.4517O}.

After extending the iSW-formulae to coupled cosmological models in Sect.~\ref{sect_homogeneous}, I compute the power spectrum of the iSW-effect in Sect.~\ref{sect_isw} for a phenomenological model in which dark matter decays into dark energy. A summary of my results is compiled in Sect.~\ref{sect_summary}. I consider spatially flat homogeneous dark energy cosmologies with adiabatic initial conditions in the cold dark matter field. Specific parameter choices are $H_0=100h \:\mathrm{km}/s/\mathrm{Mpc}$ with $h=0.72$, $\Omega_m=0.25$, $\sigma_8=0.8$ and $n_s=1$.

\section{Cosmology with coupled dark fluids}\label{sect_homogeneous}

\subsection{Decay of dark matter into dark energy}
As an example, I compute the iSW-spectra for the coupled model proposed by \citet{2008arXiv0801.1565B}, in which cold dark matter (CDM) decays into dark energy: The evolution of the dark matter density $\rho_m$ and the dark energy density $\rho_\phi$ (with an equation of state parameter $w$) is described by the system of differential equations,
\begin{equation}
\partial_t\left(\begin{array}{c}\rho_m\\ \rho_\phi\end{array}\right)
+\left(\begin{array}{cc}3H+\Gamma & 0 \\ -\Gamma & 3H(1+w)\end{array}\right)
\left(\begin{array}{c}\rho_m\\ \rho_\phi\end{array}\right) = 0,
\label{eqn_cosmology}
\end{equation}
with the cosmic time $t$ as the time variable. The Friedmann constraint is given by $3H^2=\rho_m+\rho_\phi$. The coupling constant $\Gamma$ corresponds to the CDM decay rate in units of the Hubble-constant $H_0$. Models with stable CDM and hence uncoupled fluids are recovered by setting $\Gamma=0$. I reformulate the time derivatives in the system of equations as derivatives with respect to the scale factor $a$ for introducing the common parameterisation for the dark energy equation of state \citep{1997PhRvD..56.4439T, 2003MNRAS.346..573L},
\begin{equation}
w(a) = w_0 + (1-a)w_a.
\end{equation}
With solutions for $\rho_m(a)$ and $\rho_\phi(a)$, the Hubble function $H(a)$ is can be obtained by integrating,
\begin{equation}
\frac{\dd H^2}{\dd a} = -\left(\rho_m(a)+\rho_\phi(a)\left[1+w(a)\right]\right),
\end{equation}
which in addition gives the definition of the critical density $\rho_\mathrm{crit}(a)=3H^2(a)/(8\pi G)$, with Newton's constant $G$. Initial conditions are defined at the present epoch, $H(a=1)=H_0$, $\Omega_m(a=1)=\Omega_m$ and $\Omega_\phi(a=1)=1-\Omega_m$. The comoving distance $\chi$ follows from the solution for $H(a)$,
\begin{equation}
\chi(a) = c\int_a^1\:\frac{\dd a}{a^2 H(a)},
\end{equation}
$c$ denoting the speed of light, and the density parameters can be defined by normalising $\Omega_m(a)=\rho_m(a)/\rho_\mathrm{crit}(a)$ and $\Omega_\phi(a)=\rho_\phi(a)/\rho_\mathrm{crit}(a)$. Table~\ref{table_models} gives an overview over the dark energy models considered in this paper, in terms of their dark energy properties $w_0$, $w_a$ and their CDM decay rate $\Gamma$. In the following I focus on spatially flat models with $\Omega_m+\Omega_\phi=1$, which is conserved in the time evolution of eqn.~(\ref{eqn_cosmology}).

\begin{table}\vspace{-0.1cm}
\begin{center}
\begin{tabular}{lccccccl}
\hline\hline
model				& $\Omega_m$	& $\sigma_8$	& $n_s$	& $w_0$			& $w_a$			& $\Gamma$ 	& CDM\\
\hline
$\Lambda$CDM			&0.25		& 0.8		& 1		& -1				& 0				& 0		& stable\\
$\Lambda_\Gamma$CDM	&0.25		& 0.8		& 1		& -1				& 0				& $\frac{1}{3}$	& decaying\\
$\phi$CDM				&0.25		& 0.8		& 1		& $-\frac{2}{3}$	& $-\frac{1}{3}$	& 0		& stable \\
$\phi_\Gamma$CDM		&0.25		& 0.8		& 1		& $-\frac{2}{3}$	& $-\frac{1}{3}$	& $\frac{1}{3}$	& decaying\\
\hline
\end{tabular}
\end{center}
\caption{Summary of the four dark energy models considered in this paper, in terms of the dark energy equation of state parameters $w_0$, $w_a$ and the CDM decay constant $\Gamma$.}
\label{table_models}
\end{table}

\subsection{Structure growth in dark energy cosmologies}
The homogeneous growth of the overdensity field, $\delta(\bmath{x},a)=D_+(a)\delta(\bmath{x},1)$ is described by the growth function $D_+(a)$, which is a solution to the differential equation \citep{1998ApJ...508..483W, 2003MNRAS.346..573L},
\begin{equation}
\frac{\dd^2}{\dd a^2}D_+
+\frac{1}{a}\left(3+\frac{\dd\ln H}{\dd\ln a}\right)\frac{\dd}{\dd a}D_+ = 
\frac{3}{2a^2}\Omega_m(a)D_+(a).
\label{eqn_growth}
\end{equation}
In the standard cold dark matter (SCDM) cosmology with $\Omega_m=1$ and $3+\dd\ln H/\dd\ln a=\frac{3}{2}$, the solution for $D_+(a)$ is simply the scale factor, $D_+(a)=a$. Eqn.~(\ref{eqn_growth}) can be applied for describing structure growth in coupled models as well, because the overdensity field, being the ratio between the density perturbation and the mean density, is not affected by e.g. CDM decay, as long as the dark energy fluid can be considered to be homogeneous and not to be influencing local structure formation. Nevertheless, coupled models influence the growth equation by the scaling of the Hubble function $H(a)$ and the time evolution of the density parameter $\Omega_m(a)$. 

Additionally, the initial conditions of the growth equation are different, as the cosmology does not have an asymptotic SCDM phase with $H\propto a^{-3/2}$ at early times. The inital conditions can be specified by assuming an asymptotic power-law growth $D_+(a)=a^\alpha$, $\alpha>0$. Substitution into eqn.~(\ref{eqn_growth}) yields a quadratic equation for $\alpha$, whose positive solution specifies the initial conditions $D_+(a_i)=a_i^\alpha$ and $\dd D_+(a_i)/\dd a=\alpha a_i^{\alpha-1}$ at some early time $a_i$ \citep[c.f.][]{lensing_paper}.

\begin{figure}
\resizebox{\hsize}{!}{\includegraphics{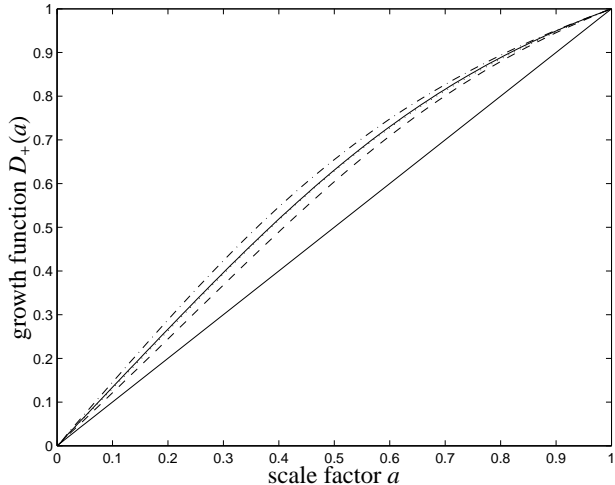}}
\caption{The growth function $D_+(a)$ as a function of scale factor $a$, for $\Lambda$CDM (solid line), $\Lambda_\Gamma$CDM with $\Gamma=\frac{1}{3}$ (dashed line), $\phi$CDM (dash-dotted line), and $\phi_\Gamma$CDM with $\Gamma=\frac{1}{3}$ (dotted line). In addition, the growth $D_+(a)=a$ for the SCDM cosmology is plotted (solid straight line).}
\label{fig_growth}
\end{figure}

Fig.~\ref{fig_growth} shows growth functions for the four exemplary cosmologies. Evolving dark energy has the property of suppressing structure formation at an earlier time, which can be partially compensated by CDM decay (because the gravitational fields generated by the overdensity $\delta(\bmath{x})$ are stronger if $\Omega_m(a)$ has a higher value), causing degeneracies between the equation of state parameters and the CDM decay rate: The $\Lambda$CDM and $\phi_\Gamma$CDM-models, for instance, are almost indistinguishable.

It is worth noting that models with decaying CDM are genuinely different from dark energy models concerning structure growth and probes which use gravitational interaction such as gravitational lensing or the iSW-effect. While it is always possible to construct an equation of state $w(a)$ for a dark energy model with stable CDM that gives the identical Hubble function as a model with decaying CDM, this degeneracy is broken in the evolution of the density parameter $\Omega_m(a)$ and the growth function $D_+(a)$ which both would be different in the two cases. Theoretically, combining observations of cosmic structure growth and the expansion history should make it to distinguish between the two families of models, which geometrical probes such as supernova observations alone would not be able to do.

\subsection{iSW-effect in coupled models}
The iSW-effect is caused by gravitational interaction of a CMB photon with a time-evolving potential $\Phi$. The fractional perturbation $\tau$ of the CMB temperature $T_\mathrm{CMB}$ is given by \citep{1967ApJ...147...73S}
\begin{equation}
\tau 
= \frac{\Delta T}{T_\mathrm{CMB}} 
\equiv \frac{2}{c^2}\int\dd\eta\: \frac{\partial\Phi}{\partial\eta} 
= -\frac{2}{c^3}\int_0^{\chi_H}\dd\chi\: a^2 H(a) \frac{\partial\Phi}{\partial a},
\label{eqn_sachs_wolfe}
\end{equation}
where $\eta$ denotes the conformal time. In the last step, I have replaced the integration variable by the comoving distance $\chi$, which is related to the conformal time by $\dd\chi = -c\dd\eta = -c\dd t/a$, and the time derivative of the growth function has been rewritten in terms of the scale factor $a$, using the definition of the Hubble function $\dd a/\dd t = aH(a)$, with the cosmic time $t$. The gravitational potential $\Phi$ follows from the Poisson equation in the comoving frame, 
\begin{equation}
\Delta\Phi = 4\pi G a^2\rho_m(a)\:\delta
\end{equation}
where Newton's constant $G$ is replaced with the critical density $\rho_\mathrm{crit}(a)=3H^2(a)/(8\pi G)$, $\rho_m(a) = \Omega_m(a)\rho_\mathrm{crit}(a)$,
\begin{equation}
\Delta\Phi = \frac{3}{2}a^2H^2(a)\Omega_m(a)\:\delta.
\end{equation}
Substitution yields a line of sight expression for the linear iSW-effect $\tau$ (using the Born-approximation and assuming linear structure formation),
\begin{equation}
\tau = 
\frac{3H_0^2}{c^3}\int_0^{\chi_H}\dd\chi\: 
a^2 H(a)\:\frac{\dd Q}{\dd a}\:\Delta^{-1}\delta,
\end{equation}
with the inverse Laplace operator $\Delta^{-1}$ solving for the potential. The function $Q(a)$ is given by
\begin{equation}
Q(a) = a^2 h^2(a)\Omega_m(a) D_+(a),
\end{equation}
with $h(a)=H(a)/H_0$, and has the derivative $\dd Q/\dd a$,
\begin{equation}
\frac{\dd Q}{\dd a} = \frac{Q}{a}
\left(2\left(1+\frac{\dd\ln H}{\dd\ln a}\right) 
+ \frac{\dd\ln\Omega_m}{\dd\ln a} + \frac{\dd\ln D_+}{\dd\ln a}\right),
\label{eqn_dq}
\end{equation}
i.e. the iSW-effect measures the sum of the scalings of the Hubble function, of the matter density and of the growth function, integrated along a line of sight. 

\begin{figure}
\resizebox{\hsize}{!}{\includegraphics{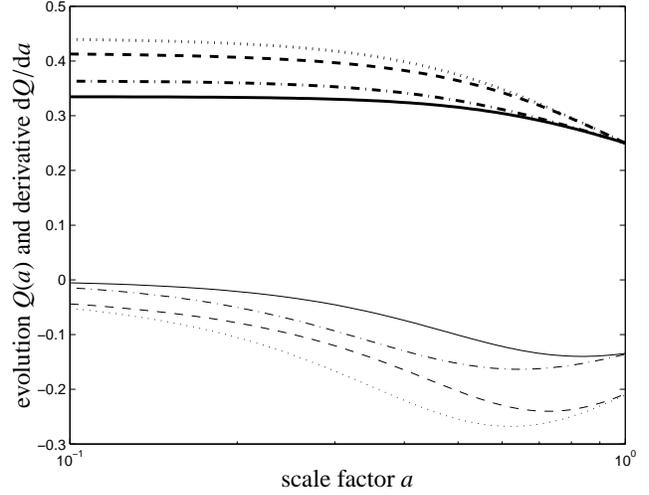}}
\caption{Time evolution $Q(a)$ of the iSW source term (thick lines) and its derivative $\dd Q(a)/\dd a$ (thin lines) as functions of scale factor $a$, for $\Lambda$CDM (solid line), $\Lambda_\Gamma$CDM with $\Gamma=\frac{1}{3}$ (dashed line), $\phi$CDM (dash-dotted line), and $\phi_\Gamma$CDM with $\Gamma=\frac{1}{3}$ (dotted line).}
\label{fig_qval}
\end{figure}

The time evolution of the iSW-source term $Q(a)$ and its derivative $\dd Q/\dd a$ are depicted in Fig.~\ref{fig_qval}. All models intersect at $Q(a)=\Omega_m$ at the current epoch, and models with evolving dark energy or CDM decay show higher values for $Q(a)$ in the past. Both the models with decay and the models with evolving dark energy assume similar values for $\dd Q/\dd a$ at low and high redshifts, respectively, and evolving dark energy shifts the curves horizontally. Again, the model with evolving dark energy and CDM decay produces the highest amplitudes for the derivative $\dd Q/\dd a$. 

Physically, the expression for $Q(a)$ corresponds to the three ways in which models with CDM decay generate higher amplitudes for the iSW-effect, compared to models with stable CDM: Apart from the growth factor $D_+(a)$ and the Hubble function $H(a)$ (due to the time derivative of $D_+$), the matter density has had higher values in the past, and hence the fluctuations in the gravitational potentials generated by the overdensity field $\delta$ are stronger, which is described by $\Omega_m(a)$. Further interesting points include:
\begin{itemize}
\item{The derivative $\dd Q/\dd a$ in fact vanishes in the SCDM cosmology with $\Omega_m(a)=1$, $D_+(a)=a$ and $h(a)=a^{-3/2}$, because $Q(a)\equiv 1$ in this case. It is perhaps even more illustrative to look directly at the expression for $\dd Q/\dd a$ (eqn.~\ref{eqn_dq}) which is zero in SCDM cosmologies because (i) the logarithmic derivative of the Hubble function $H(a)$ is equal $-3/2$ (reflecting the matter domination), (ii) the logarithmic derivative of the matter density parameter vanishes (due to the stability of CDM), and the logarithmic derivative of the growth function $D_+(a)$ is equal to one (a combination of the two). From this point of view, both the growth proportional to $a$ and stable CDM are the necessary conditions that make the linear iSW-effect disappear in Einstein-de Sitter universes.}
\item{The two cosmologies $\Lambda$CDM and $\phi_\Gamma$CDM, which are almost degenerate in terms of their growth rates, give rise to very different $Q(a)$ and $\dd Q/\dd a$ functions, which is a result of the their dependence on $H(a)$ and $\Omega_m(a)$, and hence to a different iSW-effect.}
\item{In models with no coupling between the dark fluids, the matter density decreases $\rho_m(a)\propto a^{-3}$, and 
\begin{equation}
\frac{\Omega_m(a)}{\Omega_m} = \frac{H_0^2}{a^3H^2(a)}
\end{equation}
applies. This allows the reformulation of the comoving Poisson equation in the familiar form,
\begin{equation}
\Delta\Phi = \frac{3H_0^2\Omega_m}{2a}\delta.
\end{equation}
Substitution into the line of sight integral yields the familiar equation for the iSW-effect in spatially flat dark energy cosmologies:
\begin{equation}
\tau = 
\frac{3H_0^2\Omega_m}{c^3}
\int_0^{\chi_H}\dd\chi\:a^2H(a)\left(\frac{\dd}{\dd a}\frac{D_+}{a}\right)\Delta^{-1}\delta,
\end{equation}
and $Q(a)$ reduces to $\Omega_m D_+(a)/a$ with a constant $\Omega_m$.
}
\item{An analogous consideration applies to gravitational lensing: The weak lensing convergence $\kappa$ in coupled models reads 
\begin{equation}
\kappa = \frac{3}{2c^2}\int_0^{\chi_H}\dd\chi\: a^2 H^2(a) \Omega_m(a)\: G(\chi)\chi\:D_+\delta,
\end{equation}
which replaces the common expression \citep{1992grle.book.....S, 2001PhR...340..291B},
\begin{equation}
\kappa = \frac{3H_0^2\Omega_m}{2c^2}\int_0^{\chi_H}\dd\chi\: G(\chi)\chi\:\frac{D_+}{a}\delta.
\end{equation}
$G(\chi)$ is the lensing efficiency function of the background galaxy sample, which itself is influenced by the Hubble function $H(a)$.}
\end{itemize}
The derivation above holds for general coupled cosmologies and is not restricted to the case of CDM decay, for which I exemplarily compute the iSW-spectra in the next chapter.

\section{iSW angular power spectra}\label{sect_isw}
In summary, the line of sight integrals for the iSW-temperature perturbation $\tau$ and the galaxy density $\gamma$ read:
\begin{eqnarray}
\tau & = & \frac{3}{c}\int_0^{\chi_H}\dd\chi\: a^2H(a)\frac{\dd Q}{\dd a}\varphi,\\
\gamma & = & b\int_0^{\chi_H}\dd\chi\: p(z)\frac{\dd z}{\dd\chi} D_+(\chi)\delta,
\end{eqnarray}
where I have defined the dimensionless potential $\varphi\equiv\Delta^{-1}\delta/d_H^2$, rescaled with the square of the Hubble distance $d_H=c/H_0$ for convenience. The weighting functions
\begin{eqnarray}
W_\tau(\chi) & = & \frac{3}{c}a^2 H(a) \frac{\dd Q}{\dd a},\\
W_\gamma(\chi) & = & p(z)\frac{\dd z}{\dd\chi} b D_+(a),
\end{eqnarray}
can be identified, which allow the expressions for the iSW-auto spectrum $C_{\tau\tau}(\ell)$, the iSW-cross spectrum $C_{\tau\gamma}(\ell)$ and the galaxy spectrum $C_{\gamma\gamma(\ell)}$ to be written in a compact notation, applying a Limber-projection \citep{1954ApJ...119..655L} in the flat-sky approximation, for simplicity:
\begin{eqnarray}
C_{\tau\tau}(\ell) & = & \int_0^{\chi_H}\dd\chi\: \frac{W_\tau^2(\chi)}{\chi^2}P_{\varphi\varphi}(k=\ell/\chi),\\
C_{\tau\gamma}(\ell) & = & \int_0^{\chi_H}\dd\chi\: \frac{W_\tau(\chi)W_\gamma(\chi)}{\chi^2}P_{\varphi\delta}(k=\ell/\chi),\\
C_{\gamma\gamma}(\ell) & = & \int_0^{\chi_H}\dd\chi\: \frac{W_\gamma^2(\chi)}{\chi^2}P_{\delta\delta}(k=\ell/\chi),
\end{eqnarray}
with the cross-spectrum $P_{\varphi\delta}(k) = P_{\delta\delta}(k) / (d_H k)^2$ and the spectrum $P_{\varphi\varphi}(k) = P_{\delta\delta}(k) / (d_H k)^4$ of the potential $\varphi$. 

For the CDM power spectrum I make the ansatz $P(k)\propto k^{n_s} T^2(k)$, with the transfer function \citep{1986ApJ...304...15B},
\begin{displaymath}
T(q) = \frac{\ln(1+2.34q)}{2.34q}\left(1+3.89q+(16.1q)^2+(5.46q)^3+(6.71q)^4\right)^{-\frac{1}{4}},
\end{displaymath}
where the wave vector $q$ is given in units of the shape parameter $\Omega_m h$. $P(k)$ is normalised to the value $\sigma_8$ on the scale $R=8~\mathrm{Mpc}/h$,
\begin{equation}
\sigma_R^2 = \frac{1}{2\pi^2}\int\dd k\: k^2 W^2(kR) P(k),
\end{equation}
with a Fourier-transformed spherical top-hat $W(x)=3j_1(x)/x$ as the filter function. $j_\ell(x)$ denotes the spherical Bessel function of the first kind of order $\ell$ \citep{1972hmf..book.....A}. The contribution of nonlinear structure formation to the power spectrum $P(k)$ was described with the model proposed by \citet{2003MNRAS.341.1311S}, yielding the power spectrum $P_{\delta\delta}(k)$, where the parameterisation of the nonlinear contribution to the fluctuation amplitude with $\Omega_m(a)$ as the time variable is suited for coupled models.

I use the redshift distribution of the main galaxy sample of the Dark UNiverse Explorer\footnote{\tt http://www.dune-mission.net/} \citep{2008arXiv0802.2522R}, which will observe half of the sky out to redshifts of unity and which will be of particular use for iSW-observations \citep{2008arXiv0802.0983D}:
\begin{equation}
p(z)\dd z = p_0\left(\frac{z}{z_0}\right)^2\exp\left(-\left(\frac{z}{z_0}\right)^\beta\right)\dd z
\quad\mathrm{with}\quad \frac{1}{p_0}=\frac{z_0}{\beta}\Gamma\left(\frac{3}{\beta}\right),
\end{equation}
with $\beta=3/2$ and $z_0=0.64$, which results in a median redshift of $z_\mathrm{med}=0.9$. For simplicity, the bias $b$ is assumed to be non-evolving and equal to unity. 

\begin{figure}
\resizebox{\hsize}{!}{\includegraphics{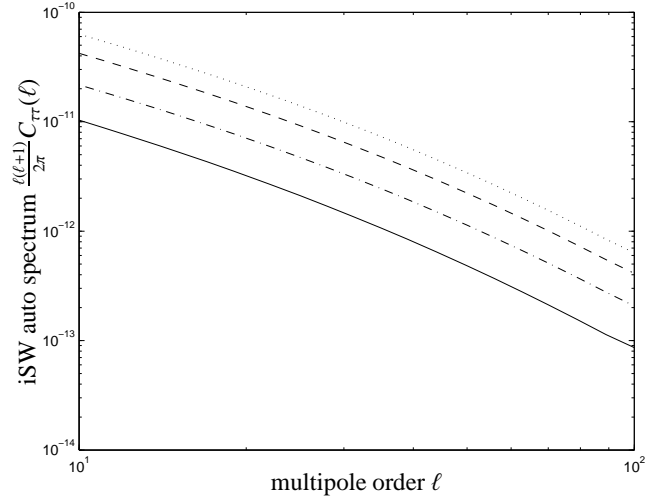}}
\caption{iSW auto power spectra $C_{\tau\tau}(\ell)$ as a function of inverse angular scale $\ell$, for $\Lambda$CDM (solid line), $\Lambda_\Gamma$CDM with $\Gamma=\frac{1}{3}$ (dashed line), $\phi$CDM (dash-dotted line), and $\phi_\Gamma$CDM with $\Gamma=\frac{1}{3}$ (dotted line).}
\label{fig_isw_auto}
\end{figure}

\begin{figure}
\resizebox{\hsize}{!}{\includegraphics{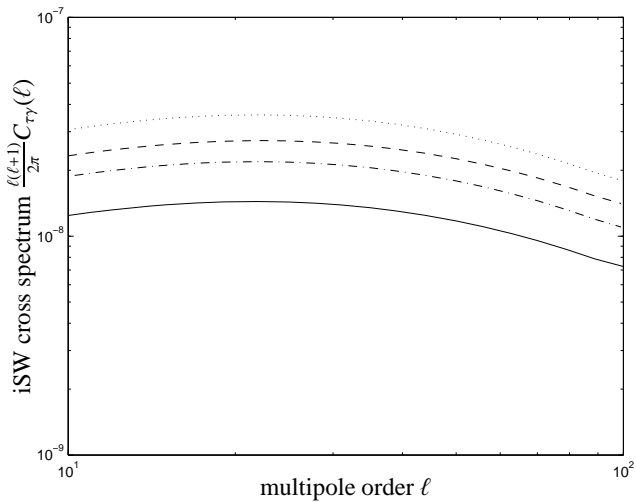}}
\caption{iSW cross power spectra $C_{\tau\gamma}(\ell)$ as a function of inverse angular scale $\ell$, for $\Lambda$CDM (solid line), $\Lambda_\Gamma$CDM with $\Gamma=\frac{1}{3}$ (dashed line), $\phi$CDM (dash-dotted line), and $\phi_\Gamma$CDM with $\Gamma=\frac{1}{3}$ (dotted line).}
\label{fig_isw_cross}
\end{figure}

The angular spectra $C_{\tau\tau}(\ell)$ and $C_{\tau\gamma}(\ell)$ are shown in Figs.~\ref{fig_isw_auto} and~\ref{fig_isw_cross}, respectively, for the decaying CDM cosmology outlined in Sect.~\ref{sect_homogeneous}. Differences between the cosmologies considered are not strongly scale dependent as expected for changes on the homogeneous level and observing linear structure formation, and amount to half an order of magnitude in the auto spectrum $C_{\tau\tau}(\ell)$. The cross spectrum $C_{\tau\gamma}(\ell)$ exhibits differences up to a factor of two. Both spectra seem to be equally sensitive to the dark energy equation of state as well as to the coupling term $\Gamma$, and these two parameters are naturally degenerate with the fluctuation amplitude $\sigma_8$ of the density field. For the chosen redshift distribution, most of the cross signal originates at a redshift of $z\simeq1.0$ ($a=0.5$, c.f. Fig.~\ref{fig_qval}), where the source function $\dd Q/\dd a$ is significantly stronger in models with either evolving dark energy or CDM decay, compared to $\Lambda$CDM. The spectra show degeneracies between $w$, $\Gamma$ and $\sigma_8$, such that a precision measurement will have to rely on a good prior on $\sigma_8$ in order to unlock the sensitivity of the iSW-effect on $w$ and $\Gamma$.

\section{Summary}\label{sect_summary}
In this paper, I extend the expressions for the iSW-effect to cosmological models with couplings between the dark matter and dark energy densities, and compute the auto and cross correlation angular spectra for a coupled cosmology, in which dark matter decays into dark energy.

\begin{itemize}
\item{The iSW-effect reflects couplings between the dark matter and dark energy fluids by three mechanisms: The Hubble function $H(a)$ shows a different scaling in the past, the growth equation has different initial conditions and a different time evolution, and the Poisson equation, which relates the gravitational potential to the overdensity field, is affected by the evolution of the matter density $\Omega_m(a)$. The iSW-formula reduces to the familiar form if $\Omega_m(a)/\Omega_m=H_0^2/H^2(a)/a^3$ is substituted, which holds in uncoupled models.}
\item{The iSW-spectrum $C_{\tau\tau}(\ell)$ shows higher amplitudes in models with couplings as well as in models with evolving dark energy, compared to models with stable matter or those with a cosmological constant $\Lambda$. For the (extreme) choice of $\Gamma$ and $w$ the cross-spectrum $C_{\tau\gamma}(\ell)$ assumes amplitudes which are twice as large in dark energy models with evolving $w$ or in coupled models, i.e. the sensitivity of the iSW-effect towards the mean dark energy equation of state parameter and coupling terms is comparable. The shape of the spectra suggests degeneracies between $w$, $\Gamma$ and $\sigma_8$, which all increase the overall power. For that reason, good independent priors on $\sigma_8$ are essential to constraints on $w$ and $\Gamma$ from the iSW-effect.}
\item{Because of the weakness of the iSW-effect it will be difficult derive strong constraints on size of the interaction term, or the specific type of the coupling. Gravitational lensing, on the contrary, is equally affected by a violation of the scaling of the matter density, and is likely to yield constraints comparable to those on the dark energy equation of state parameters, possibly to an accuracy of a few percent with upcoming surveys \citep{2008arXiv0803.1640L}. A companion paper considers structure formation in coupled models and examins the measurement of the CDM decay rate $\Gamma$ with weak lensing bispectrum tomography \citep{lensing_paper}.}
\end{itemize}

\section*{Acknowledgements}
I would like to thank R. Maartens and G.A. Caldera Cabral for sharing their expertise cosmologies with coupled dark matter and dark energy, and R. Maartens in particular for valuable comments on the paper. I am grateful to M. Douspis and A. Refregier for giving me the DUNE galaxy redshift distribution.

\bibliography{bibtex/aamnem,bibtex/references}

\begin{thebibliography}{}

\bibitem[\protect\citeauthoryear{{Abramowitz} \& {Stegun}}{{Abramowitz} \&
  {Stegun}}{1972}]{1972hmf..book.....A}
{Abramowitz} M.,  {Stegun} I.~A.,  1972, {Handbook of Mathematical Functions}.
Handbook of Mathematical Functions, New York: Dover, 1972

\bibitem[\protect\citeauthoryear{{Bardeen}, {Bond}, {Kaiser} \&
  {Szalay}}{{Bardeen} et~al.}{1986}]{1986ApJ...304...15B}
{Bardeen} J.~M.,  {Bond} J.~R.,  {Kaiser} N.,    {Szalay} A.~S.,  1986, \apj,
  304, 15

\bibitem[\protect\citeauthoryear{{Bartelmann} \& {Schneider}}{{Bartelmann} \&
  {Schneider}}{2001}]{2001PhR...340..291B}
{Bartelmann} M.,  {Schneider} P.,  2001, \physrep, 340, 291

\bibitem[\protect\citeauthoryear{{Boehmer}, {Caldera-Cabral}, {Lazkoz} \&
  {Maartens}}{{Boehmer} et~al.}{2008}]{2008arXiv0801.1565B}
{Boehmer} C.~G.,  {Caldera-Cabral} G.,  {Lazkoz} R.,    {Maartens} R.,  2008,
  ArXiv 0801.1565, 801

\bibitem[\protect\citeauthoryear{{Boughn} \& {Crittenden}}{{Boughn} \&
  {Crittenden}}{2003}]{2003AIPC..666...67B}
{Boughn} S.~P.,  {Crittenden} R.~G.,  2003, in {Holt} S.~H.,  {Reynolds} C.~S.,
   eds, The Emergence of Cosmic Structure Vol.~666 of American Institute of
  Physics Conference Series, {The Absence of the Integrated Sachs-Wolfe Effect:
  Constraints on a Cosmological Constant}.
pp 67--70

\bibitem[\protect\citeauthoryear{{Chimento}, {Jakubi}, {Pav{\'o}n} \&
  {Zimdahl}}{{Chimento} et~al.}{2003}]{2003PhRvD..67h3513C}
{Chimento} L.~P.,  {Jakubi} A.~S.,  {Pav{\'o}n} D.,    {Zimdahl} W.,  2003,
  \prd, 67, 083513

\bibitem[\protect\citeauthoryear{{Cooray}}{{Cooray}}{2002}]{2002PhRvD..65h3518%
C}
{Cooray} A.,  2002, \prd, 65, 083518

\bibitem[\protect\citeauthoryear{{Crittenden} \& {Turok}}{{Crittenden} \&
  {Turok}}{1996}]{1996PhRvL..76..575C}
{Crittenden} R.~G.,  {Turok} N.,  1996, Physical Review Letters, 76, 575

\bibitem[\protect\citeauthoryear{{Douspis}, {Castro}, {Caprini} \&
  {Aghanim}}{{Douspis} et~al.}{2008}]{2008arXiv0802.0983D}
{Douspis} M.,  {Castro} P.~G.,  {Caprini} C.,    {Aghanim} N.,  2008, ArXiv
  0802.0983, 802

\bibitem[\protect\citeauthoryear{{Fosalba}, {Gazta{\~ n}aga} \&
  {Castander}}{{Fosalba} et~al.}{2003}]{2003ApJ...597L..89F}
{Fosalba} P.,  {Gazta{\~ n}aga} E.,    {Castander} F.~J.,  2003, \apjl, 597,
  L89

\bibitem[\protect\citeauthoryear{{Giannantonio}, {Crittenden}, {Nichol},
  {Scranton}, {Richards}, {Myers}, {Brunner}, {Gray}, {Connolly} \&
  {Schneider}}{{Giannantonio} et~al.}{2006}]{2006PhRvD..74f3520G}
{Giannantonio} T.,  {Crittenden} R.~G.,  {Nichol} R.~C.,  {Scranton} R.,
  {Richards} G.~T.,  {Myers} A.~D.,  {Brunner} R.~J.,  {Gray} A.~G.,
  {Connolly} A.~J.,    {Schneider} D.~P.,  2006, \prd, 74, 063520

\bibitem[\protect\citeauthoryear{{Giannantonio}, {Scranton}, {Crittenden},
  {Nichol}, {Boughn}, {Myers} \& {Richards}}{{Giannantonio}
  et~al.}{2008}]{2008arXiv0801.4380G}
{Giannantonio} T.,  {Scranton} R.,  {Crittenden} R.~G.,  {Nichol} R.~C.,
  {Boughn} S.~P.,  {Myers} A.~D.,    {Richards} G.~T.,  2008, ArXiv 0801.4380,
  801

\bibitem[\protect\citeauthoryear{{He} \& {Wang}}{{He} \&
  {Wang}}{2008}]{2008arXiv0801.4233H}
{He} J.-H.,  {Wang} B.,  2008, ArXiv 0801.4233, 801

\bibitem[\protect\citeauthoryear{{Hu} \& {Sugiyama}}{{Hu} \&
  {Sugiyama}}{1994}]{1994PhRvD..50..627H}
{Hu} W.,  {Sugiyama} N.,  1994, \prd, 50, 627

\bibitem[\protect\citeauthoryear{{La Vacca} \& {Colombo}}{{La Vacca} \&
  {Colombo}}{2008}]{2008arXiv0803.1640L}
{La Vacca} G.,  {Colombo} L.~P.~L.,  2008, ArXiv 0803.1640, 803

\bibitem[\protect\citeauthoryear{{Limber}}{{Limber}}{1954}]{1954ApJ...119..655%
L}
{Limber} D.~N.,  1954, \apj, 119, 655

\bibitem[\protect\citeauthoryear{{Linder} \& {Jenkins}}{{Linder} \&
  {Jenkins}}{2003}]{2003MNRAS.346..573L}
{Linder} E.~V.,  {Jenkins} A.,  2003, \mnras, 346, 573

\bibitem[\protect\citeauthoryear{{Olivares}, {Atrio-Barandela} \&
  {Pavon}}{{Olivares} et~al.}{2007}]{2007arXiv0706.3860O}
{Olivares} G.,  {Atrio-Barandela} F.,    {Pavon} D.,  2007, ArXiv 0706.3860,
  706

\bibitem[\protect\citeauthoryear{{Olivares}, {Atrio-Brandela} \&
  {Pavon}}{{Olivares} et~al.}{2008}]{2008arXiv0801.4517O}
{Olivares} G.,  {Atrio-Brandela} F.,    {Pavon} D.,  2008, ArXiv 0801.4517, 801

\bibitem[\protect\citeauthoryear{{Pettorino} \& {Baccigalupi}}{{Pettorino} \&
  {Baccigalupi}}{2008}]{2008arXiv0802.1086P}
{Pettorino} V.,  {Baccigalupi} C.,  2008, ArXiv 0802.1086, 802

\bibitem[\protect\citeauthoryear{{Rassat}, {Land}, {Lahav} \&
  {Abdalla}}{{Rassat} et~al.}{2007}]{2007MNRAS.377.1085R}
{Rassat} A.,  {Land} K.,  {Lahav} O.,    {Abdalla} F.~B.,  2007, \mnras, 377,
  1085

\bibitem[\protect\citeauthoryear{{Rees} \& {Sciama}}{{Rees} \&
  {Sciama}}{1968}]{rees_sciama_orig}
{Rees} M.~J.,  {Sciama} D.~W.,  1968, Nature, 217, 511

\bibitem[\protect\citeauthoryear{{Refregier} \& {the DUNE
  collaboration}}{{Refregier} \& {the DUNE
  collaboration}}{2008}]{2008arXiv0802.2522R}
{Refregier} A.,  {the DUNE collaboration} 2008, ArXiv 0802.2522, 802

\bibitem[\protect\citeauthoryear{{Sachs} \& {Wolfe}}{{Sachs} \&
  {Wolfe}}{1967}]{1967ApJ...147...73S}
{Sachs} R.~K.,  {Wolfe} A.~M.,  1967, \apj, 147, 73

\bibitem[\protect\citeauthoryear{{Sch{\"a}fer} \& {Bartelmann}}{{Sch{\"a}fer}
  \& {Bartelmann}}{2006}]{2006MNRAS.369..425S}
{Sch{\"a}fer} B.~M.,  {Bartelmann} M.,  2006, \mnras, 369, 425

\bibitem[\protect\citeauthoryear{{Sch{\"a}fer}, {Caldera Cabral} \&
  {Maartens}}{{Sch{\"a}fer} et~al.}{2008}]{lensing_paper}
{Sch{\"a}fer} B.~M.,  {Caldera Cabral} G.~A.,    {Maartens} R.,  2008, ArXiv
  0803.2232, 803

\bibitem[\protect\citeauthoryear{{Schneider}, {Ehlers} \& {Falco}}{{Schneider}
  et~al.}{1992}]{1992grle.book.....S}
{Schneider} P.,  {Ehlers} J.,    {Falco} E.~E.,  1992, {Gravitational Lenses}.
Springer-Verlag Berlin Heidelberg New York.

\bibitem[\protect\citeauthoryear{{Smith}, {Peacock}, {Jenkins}, {White},
  {Frenk}, {Pearce}, {Thomas}, {Efstathiou} \& {Couchman}}{{Smith}
  et~al.}{2003}]{2003MNRAS.341.1311S}
{Smith} R.~E.,  {Peacock} J.~A.,  {Jenkins} A.,  {White} S.~D.~M.,  {Frenk}
  C.~S.,  {Pearce} F.~R.,  {Thomas} P.~A.,  {Efstathiou} G.,    {Couchman}
  H.~M.~P.,  2003, \mnras, 341, 1311

\bibitem[\protect\citeauthoryear{{Turner} \& {White}}{{Turner} \&
  {White}}{1997}]{1997PhRvD..56.4439T}
{Turner} M.~S.,  {White} M.,  1997, \prd, 56, 4439

\bibitem[\protect\citeauthoryear{{Wang} \& {Steinhardt}}{{Wang} \&
  {Steinhardt}}{1998}]{1998ApJ...508..483W}
{Wang} L.,  {Steinhardt} P.~J.,  1998, \apj, 508, 483

\end{thebibliography}
\bibliographystyle{mn2e}

\bsp

\label{lastpage}

\end{document}